\documentclass[letterpaper, 10 pt, conference]{ieeeconf}  
\IEEEoverridecommandlockouts                             
\usepackage{graphics} % for pdf, bitmapped graphics files
\usepackage{epsfig} % for postscript graphics files
\usepackage{times} % assumes new font selection scheme installed
\usepackage{amsmath} % assumes amsmath package installed
\usepackage{amssymb}  % assumes amsmath package installed
\usepackage{xcolor}
\usepackage{bm}
\usepackage{cancel}
\usepackage{color}
\usepackage{setspace}
\usepackage{wrapfig}
\usepackage{tikz}
\usetikzlibrary{shapes,arrows.meta}
\usetikzlibrary{positioning}
\usepackage{caption}
\usepackage{subcaption}
\usepackage{booktabs}
\usepackage{multirow}
\usepackage{multicol}
\usepackage{soul}
\usepackage{cite}
\usepackage[normalem]{ulem}
\usepackage{xcolor}

\newcommand{\ellone}{$\mathcal{L}_1$ }
\let\labelindent\relax % Use for IEEE templates to suppress the already defined \labelindent
\usepackage{enumitem}
\usepackage{tabularx}
\definecolor{lcolor}{rgb}{0.79,0.88,1.0}
\definecolor{bcolor}{RGB}{252,184,203}
\definecolor{ccolor}{RGB}{234,208,255}
\tikzstyle{sqblock} = [draw, fill=bcolor!20, rectangle, 
    minimum height=2em, minimum width=3em]
\tikzstyle{block} = [draw, fill=bcolor!20, rectangle, 
    minimum height=2em, minimum width=3em]
\tikzstyle{sum} = [draw, fill=bcolor!20, circle, node distance=1cm]
\tikzstyle{input} = [coordinate]
\tikzstyle{output} = [coordinate]
\tikzstyle{phantom} = [coordinate]

\newcommand{\norm}[1]{\lVert #1 \rVert}
\newcommand{\dd}{\text{d}}
\newcommand{\diag}[1]{\text{diag}[#1]}

\title{\LARGE \bf
$\mathcal{L}_1$ Adaptive Augmentation for Geometric Tracking Control of Quadrotors% on $SE(3)$
}
\author{Zhuohuan Wu, Sheng Cheng, Kasey A. Ackerman, Aditya Gahlawat, Arun Lakshmanan, \\Pan Zhao and Naira Hovakimyan
\thanks{*This work is supported by National Aeronautics and Space Administration (NASA), Air Force Office of Scientific Research (AFOSR), National Science Foundation (NSF) Cyber Physical Systems (CPS) award \#~1932529, NSF National Robotics Initiative 2.0 (NRI-2.0) award \#~1830639, and NSF-AoF Robust Intelligence award \#~2133656.}
\thanks{Mechanical Science and Engineering, University of Illinois Urbana-Champaign, USA.
        {\tt\small \{zw24, chengs, kaacker2, gahlawat, lakshma2, panzhao2, nhovakim\}@illinois.edu}}
}

\begin{document}

\maketitle
%% for editing
% \thispagestyle{plain}
% \pagestyle{plain}
%% for submission
\thispagestyle{empty}
\pagestyle{empty}

% TODO list
% 1. update the blockdiagram: baseline uses feedback, which is not reflected
% 2. correct some formulas in the differential flatness part.
% 3. Check equation 7, 8, 9. I think if we replace $u_b$ in equation 7 and 8 with $u$, we do not need to write out equation 9 anymore. 
% 4. Check line 311, should be prediction error dynamics. 

%%%%%%%%%%%%%%%%%%%%%%%%%%%%%%%%%%%%%%%%%%%%%%%%%%%%%%%%%%%%%%%%%%%%%%%%%%%%%
\begin{abstract}
This paper introduces an $\mathcal{L}_1$ adaptive control augmentation for geometric tracking control of quadrotors. In the proposed design, the $\mathcal{L}_1$ augmentation handles nonlinear (time- and state-dependent) uncertainties in the quadrotor dynamics without assuming or enforcing parametric structures, while the baseline geometric controller achieves stabilization of the known nonlinear model of the system dynamics. The $\mathcal{L}_1$ augmentation applies to both the rotational and the translational dynamics. Experimental results demonstrate that the augmented geometric controller shows consistent and (on average five times) smaller trajectory tracking errors compared with the geometric controller alone when tested for different trajectories and under various types of uncertainties/disturbances.

% This paper introduces an $\mathcal{L}_1$ adaptive augmentation for a geometric controller and its flight tests for a quadrotor platform in a vicon lab. The $\mathcal{L}_1$ augmentation handles the nonlinear uncertainties in the quadrotor dynamics without assuming/enforcing parametric structure, while the baseline geometric controller achieves stabilization of the known nonlinear model of the system dynamics. The augmentation applies to both the rotational and the translational dynamics. We demonstrate the proposed controller design on a quadrotor in experiments. The $\mathcal{L}_1$ augmentation shows consistent smaller trajectory tracking errors compared with the geometric controller alone when tested for different paths and under various types of uncertainties/disturbances.
\end{abstract}

\begin{center}
    SUPPLEMENTARY MATERIAL
\end{center}
Video: https://youtu.be/25Z7iAkZ5xw \\
Code: https://github.com/HovakimyanResearch/L1-Mambo

%%%%%%%%%%%%%%%%%%%%%%%%%%%%%%%%%%%%%%%%%%%%%%%%%%%%%%%%%%%%%%%%%%%%%%%%%%%
\section{INTRODUCTION}
In recent years, unmanned aerial vehicles have seen an increased use across a wide range of applications, e.g., surveillance~\cite{yang2009implementation}, search and rescue~\cite{tian2020search}, agriculture~\cite{valente2013near}, and logistics~\cite{mellinger2013cooperative}. The quadrotor platform, in particular, has garnered interests due to its low-cost, relatively simple mechanical structure, and dynamic capabilities~\cite{mellinger2012trajectory,tayebi2006attitude,bhat2000topological,mayhew2011quaternion,mellinger2011minimum,lee2010geometric}. Controller synthesis for quadrotors is a challenging problem due to the unstable and underactuated nature of the dynamics. The challenges are exacerbated further due to uncertainties and disturbances, potentially leading to a loss of predictability and even stability. 

\noindent \textbf{Related Work:} Control theoretic methodologies like robust and adaptive control are used to safely operate systems subject to uncertainties. Therefore, such methodologies have been applied for quadrotor trajectory tracking in the presence of uncertainties, e.g., see the survey~\cite{emran2018review}. However, the synthesis of such controllers relies on simplifying assumptions like linear known (nominal) models and a priori known structured parametric uncertainties that can lead to overly conservative operation. For example, adaptive controllers designed in~\cite{dydek2012adaptive,selfridge2014multivariable} utilize linearized dynamics. Since the linear models are only valid around the hover-position, the quadrotor cannot track aggressive trajectories. To account for the nonlinear nature of the dynamics, the authors of~\cite{antonelli2017adaptive,cabecinhas2014nonlinear} propose adaptive controllers that use the nonlinear dynamics of the quadrotor. However, the adaptive schemes are developed under the assumption of static/constant uncertainties---an assumption that does not reflect the true time-varying and state-dependent nature of the uncertainties.
Other approaches relying on backstepping and sliding-mode techniques have also been developed, e.g.,~\cite{bouabdallah2005backstepping}. However, the switching nature of the sliding-mode control causes chattering problems that may excite high-frequency unmodelled dynamics. The chattering behavior is addressed in~\cite{labbadi2019robust}, where the controller is based on Euler angles that suffer from singularities preventing aggressive rotational maneuvers. Disturbance observer-based methods, presented in~\cite{ahmed2020disturbance,castillo2019disturbance}, require decoupled quadrotor dynamics in roll, pitch, and yaw (i.e., diagonal inertia matrix) and can only handle time-dependent and parametric uncertainties.
A geometric adaptive tracking controller is presented in~\cite{goodarzi2015geometric}, where the controller is developed directly on $SE(3)$; however, it assumes an a priori known structure of parametric uncertainties. 
 
Machine learning (ML) tools like deep neural networks (DNNs) and Gaussian process regression (GPR) can accurately learn functions using only the input-output data and do not require any a priori knowledge of the parametric structure. Thus, ML-based control methodologies have been investigated in the literature for quadrotors. In~\cite{shi2020neural}, the authors use a DNN to learn the high-order multi-vehicle aerodynamic effects. A tracking controller uses the learned model and is demonstrated in experiments for a quadrotor swarm with individual agents operating in close proximity to each other. In~\cite{bouffard2012learning}, a learning-based model predictive control (MPC) method is proposed, which uses linearized quadrotor dynamics. The authors of~\cite{torrente2021data} model aerodynamic effects using GPR, which are then used by an MPC algorithm to achieve improved tracking performance for high-velocity quadrotor operation.
Although ML-based methods can accurately approximate the unmodelled factors, the quadrotor dynamics or the flying environment can significantly change during operation. For example, uncertainties can appear in the form of wind, payload sloshing, and degradation of system capabilities. ML-based methods might fail to react to such changes when these changes are far beyond the scenarios represented in the training datasets. Therefore, ML tools have recently been incorporated within adaptive control to empower them with more agility and flexibility. For example, the work in~\cite{joshi2020asynchronous} presents a deep model reference adaptive control architecture. The uncertainties are represented by a DNN, where the inner layers are updated slowly using the collected data. In contrast, the final layer is updated in real-time in an adaptive scheme that uses the inner layers to parameterize the uncertainty. However, the work is focused on the linearized model at the hover position, preventing its applicability to aggressive trajectories. 

In light of the considerations mentioned above, we propose a control scheme for quadrotors that uses the~\ellone adaptive control as an augmentation to compensate for the nonlinear uncertainties. The use of~\ellone adaptive control is motivated by its capability of compensating for uncertainties at an arbitrarily high rate of adaptation limited only by the hardware capabilities~\cite{hovakimyan2010L1}. A critical property of the~\ellone control enabling the fast adaptation is that the estimation is decoupled from control; thus, the arbitrarily fast adaptation does not destroy the robustness of the closed-loop system~\cite{hovakimyan2010L1}. The $\mathcal{L}_1$ adaptive control has been successfully validated on NASA's AirStar $5.5\%$ subscale generic transport aircraft model~\cite{gregory2009l1,gregory2010flight}, Calspan's Learjet~\cite{ackerman2016l1,ackerman2017evaluation}, and unmanned aerial vehicles~\cite{kaminer2010path,kaminer2012time,jafarnejadsani2017optimized}. The $\mathcal{L}_1$ adaptive controller design for quadrotors has been studied in~\cite{zuo2014augmented,huynh20141}, where Euler angles are used to model the dynamics. In~\cite{kotaru2020geometric}, the authors proposed an $\mathcal{L}_1$ adaptive control augmentation of the geometric controller to avoid the singularities in Euler-angle representation. However, this $\mathcal{L}_1$ augmentation only applies to rotational dynamics, which cannot compensate for any uncertainties in the translational dynamics. %Furthermore, its design includes the rotational kinematics that are intrinsically uncertainty-free, which introduces unnecessary complexities. % in the $\mathcal{L}_1$ augmentation since the prediction error is partially defined in $SO(3)$.}
Furthermore, the adaptive scheme we propose does not rely on projection and gradient-based optimization, which can potentially cause numerical issues for implementation on the quadrotor in the presence of large adaptation gains \cite{ioannou2014l1}. Instead, we use a piecewise-constant adaption law that requires minimal computation~\cite{hovakimyan2010L1}. We present an approach that augments a geometric controller with an $\mathcal{L}_1$ adaptive controller for both rotational and translational dynamics. The geometric controller ensures exponential stability for trajectory tracking using the nominal dynamics~\cite{lee2010geometric}. Simultaneously, the $\mathcal{L}_1$ adaptive augmentation compensates for non-parametric uncertainties (both state- and time-dependent) and provides guarantees for transient performance and robustness. An initial proof of concept, in simulation, of using~\ellone controller for control of a quadrotor can be found in~\cite{pravitra2020MPPI}. We demonstrate and validate the proposed architecture through multiple experiments performed on a real quadrotor platform. The results show the $\mathcal{L}_1$ augmentation's superior tracking performance (five times smaller tracking error on average) compared with the geometric controller alone for different trajectories and under various types of disturbances and uncertainties. 

\noindent \textbf{Statement of Contributions:} i) we provide a new architecture for \ellone control augmentation of a geometric controller for both the rotational and translational dynamics that can compensate for both time- and state-dependent uncertainties; ii) the controller is designed for ease of implementation by formulating the augmentation in the Euclidean space, and by using a piecewise-constant adaptation law that requires minimal computational resources; iii) we test and validate the control scheme empirically on a real quadrotor platform subject to various uncertainties/disturbances and tasked with following different trajectories.  
 
The remainder of the paper is organized as follows: Section~\ref{sec:dynamics and geometric controller} reviews the quadrotor dynamics and the geometric controller design. Section~\ref{sec:geometric+L1} introduces the $\mathcal{L}_1$ adaptive augmentation of the geometric controller. Section~\ref{sec:experiments} shows the experimental results on a Mambo quadrotor. Finally, Section~\ref{sec:conclusions} summarizes the paper and discusses future work. 

\section{Quadrotor dynamics and geometric controller}\label{sec:dynamics and geometric controller}

% A quadrotor has four rotors and propellers located at the vertices of a square, which generate a collective thrust normal to the plane of this square and moments induced by differential thrusts.

\begin{figure}
    \centering
    \includegraphics[width=\columnwidth]{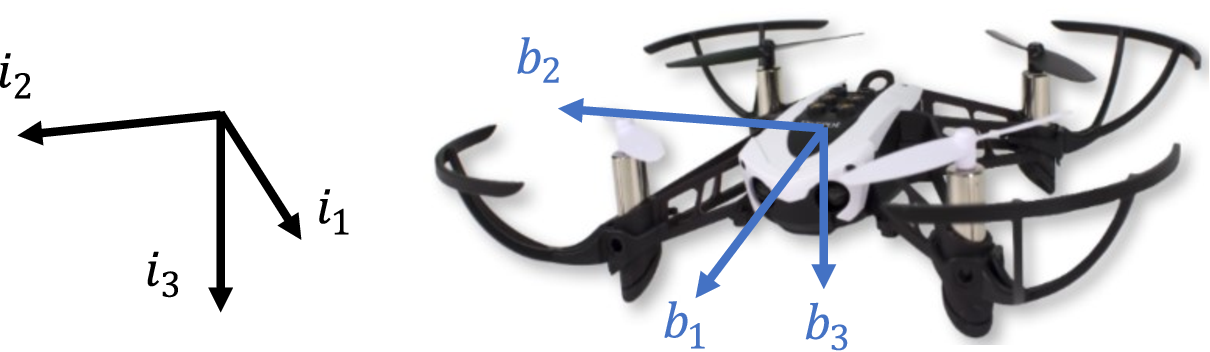}
    \caption{The Mambo quadrotor and reference frames.}
    \label{fig:mambo with frames}
    
    \vspace{-0.5cm}
\end{figure}

We choose an inertial frame and a body-fixed frame, which are spanned by unit vectors $\{ i_1,i_2,i_3 \}$ and $\{ b_1,b_2,b_3 \}$ in north-east-down directions, respectively, shown in Fig.~\ref{fig:mambo with frames}. The origin of the body-fixed frame is located at the center of mass (COM) of the quadrotor, which is assumed to be its geometric center due to its symmetric mechanical configuration. %The first and the second axes of the body-fixed frame ($b_1$ and $b_2$) lie in the plane defined by the center of the four rotors and point right and forward, respectively. The third body-fixed axis $b_3$ is normal to the plane and pointing upwards. 

The configuration space of the quadrotor is defined by the location of its COM and the attitude with respect to the inertial frame, i.e., the configuration manifold is the special Euclidean group $SE(3)$: the Cartesian product of $\mathbb{R}^3$ and the special orthogonal group $SO(3) = {\{ R \in \mathbb{R}^{3 \times 3}| R^{\top} R=I,\ \text{det} (R)=1 \}}$. By the definition of the rotation matrix $R \in SO(3)$, the direction of the $i$th body-fixed axis $b_i$ is given by $Re_i$ in the inertial frame, where $e_i$ is the unit vector with the $i$th element being 1 for $i \in \{1,2,3\}$. The rotation matrix can be obtained through the Euler angles yaw $\psi$, roll $\phi$, and pitch $\theta$ in the 3-2-1 rotation sequence~\cite{diebel2006representing}.

The equations of motion of a quadrotor~\cite{lee2010geometric} are
\begin{subequations}\label{equ:quadrotor dynamics}
\begin{align}
    \dot{p}= & v,\label{eq:position integral} \\
    \dot{v}= & ge_3-\frac{f}{m}Re_3, \label{equ:translational dynamics}\\
    \dot{R} = & R \Omega^{\wedge},\label{eq:angle integral}\\
    \dot{\Omega}= & J^{-1}(M - \Omega \times J\Omega) \label{equ:rotational dynamics},
\end{align}
\end{subequations}
where $p \in \mathbb{R}^3$ and $v \in \mathbb{R}^3$ are the position and velocity of the quadrotor's COM in the inertial frame, respectively, ${\Omega \in \mathbb{R}^3}$ is the angular velocity in the body-fixed frame, $g$ is the gravitational acceleration, $m$ is the vehicle mass, $J \in \mathbb{R}^{3 \times 3}$ is the moment of inertia matrix calculated in the body-fixed frame, $f$ is the collective thrust, and $M \in \mathbb{R}^3$ is the moment in the body-fixed frame. 
The \textit{wedge} operator $\cdot^{\wedge}:\mathbb{R}^3 \rightarrow \mathfrak{so}(3)$ denotes the mapping to the space of skew-symmetric matrices. Note that we suppress temporal dependencies to maintain clarity of exposition unless required.

We use \eqref{equ:quadrotor dynamics} as the nominal dynamics of the quadrotor to design the geometric controller. Uncertainties in the dynamics will be introduced when we discuss the $\mathcal{L}_1$ adaptive control augmentation in Section~\ref{sec:geometric+L1}. 
We do not consider the motor dynamics and the propellers' aerodynamic effects. Therefore, the thrust $f$ and moment $M$ are assumed to be linear in the squared motor speeds~\cite{mellinger2011minimum}. We choose $f$ and $M$ as the control in~\eqref{equ:quadrotor dynamics}, and they are achieved using the desired motor speeds (see \cite{mellinger2011minimum} for details in computation).

% We do not consider the motor dynamics and the propellers' aerodynamic effects. Therefore, we model the thrust $f$ and moment $M$ to be directly controlled via the motors' speed $\{\omega_i\}_{i=1}^4$, i.e., 
% \begin{equation}\label{eq:motor mapping to thrust and moment}
%     \begin{bmatrix}
%         f \\ M_1 \\ M_2 \\ M_3
%     \end{bmatrix} = 
%     \begin{bmatrix}
%         k_F & k_F & k_F & k_F \\
%         -k_F L   & -k_F L & k_F L & k_F L \\
%         k_F L & -k_F L & -k_F L & k_F L \\
%         k_M & -k_M & k_M & -k_M 
%     \end{bmatrix}
%     \begin{bmatrix}
%         \omega_1^2 \\ \omega_2^2 \\\omega_3^2 \\\omega_4^2 
%     \end{bmatrix},
% \end{equation}
% where $L$ is the distance from the center of the quadrotor to the side of the square, and $k_F$ and $k_M$ are the thrust and moment coefficients, respectively.
% We choose $f$ and $M$ as the control in~\eqref{equ:quadrotor dynamics}, and they are achieved using the desired motor speeds that are computed by inverting the linearization of \eqref{eq:motor mapping to thrust and moment} about $\omega_i = (f/4k_F)^{1/2}$~\cite{mellinger2011minimum}. % NOTE: the coefficient matrix is invertible: its determinant is -8*k_F^3*L^2*k_M, which is nonzero. But we cannot obtain the desired rotational speed by simply inversing the equation. The reason is that the domain of the coefficient matrix is R^4. Yet the expected domain is R+^4 due to the squared motor speed. Hence, there's no guarantee for omega_i^2 to be nonnegative when we use matrix inversion.

The design of the geometric controller follows \cite{lee2010geometric,mellinger2011minimum}, where the goal is to have the quadrotor follow prescribed trajectory $p_d(t) \in \mathbb{R}^3$ and yaw $\psi_d(t)$ for time $t$ in a prescribed interval $[0,t_f]$.
The quadrotor dynamics with inputs $f$ and $M$ is differentially flat with flat outputs $p_d$ and $\psi_d$ \cite{mellinger2011minimum}, meaning that the state and inputs of system \eqref{equ:quadrotor dynamics} can be expressed in terms of the flat outputs and their derivatives \cite[Sec.~2]{van1998real}.
Therefore, the underactuated quadrotor can follow a smooth trajectory in the space of flat outputs $\{p_d,\psi_d\}$. 

Define the position and velocity error as $e_p = p-p_d$ and $e_v = \dot{p}-\dot{p}_d$, respectively.
% \begin{equation}
%     e_p = p-p_d,\quad e_v = \dot{p}-\dot{p}_d.
% \end{equation}
We compute the vector of desired force $F_d \in \mathbb{R}^3$ by $F_d = -K_p e_p-K_v e_v - mg e_3 + m \ddot{p}_d$, 
% \begin{equation}
%     F_d = -K_p e_p-K_v e_v - mg e_3 + m \ddot{p}_d,
% \end{equation}
where $K_p,K_v \in \mathbb{R}^{3 \times 3}$ are user-selected positive-definite gain matrices. Projecting the desired force $F_d$ to the body-fixed $z$-axis $Re_3$, we obtain the desired thrust $f$ as the first control input:
\begin{equation}\label{eq:thrust control}
    f = -F_d \cdot (Re_3).
\end{equation}
The desired thrust is used for controlling the translational motion of the quadrotor. For controlling the rotational motion, we use the moment $M$. We first define the desired rotation matrix $R_d$ as follows. The desired $z$-axis is along the desired force, i.e., $b_{3d} = F_d/\norm{F_d}$, where we assume that $\norm{F_d} \neq 0$. Note that $b_{3d}$ is the third column of $R_d$, and we compute the other two columns using an intermediate axis $b_{\text{int}}=[\cos \psi_d \ \sin \psi_d \ 0]^\top$ associated with the desired yaw $\psi_d$.
The second and first column of the desired rotation matrix are $b_{2d} = b_{3d} \times b_{\text{int}}/\norm{b_{3d} \times b_{\text{int}}}$ and $ b_{1d} = b_{2d} \times b_{3d}$, respectively, where we assume $\norm{b_{3d} \times b_{\text{int}}}\neq 0$. We now have $R_d = [b_{1d} \ b_{2d} \ b_{3d}]$.
The attitude error is $e_R = ( R_d^\top R - R^\top R_d)^{\vee}/2$, where $^\vee$ is the \emph{vee} operator that takes elements of $\mathfrak{so}(3)$ to $\mathbb{R}^3$. The angular velocity error is $e_{\Omega} = \Omega -R^\top R_d \Omega_d$, 
% \sout{where $\Omega_d$ is computed using the differential flatness \cite[Sec.~III]{mellinger2011minimum}. Specifically,
% \begin{equation}
%     \Omega_d^\top = [\Omega_{1d} \ \Omega_{2d} \ \Omega_{3d}]=[-h_{\Omega} \cdot b_{2d}\     h_{\Omega} \cdot b_{1d}\      \dot{\psi}_d e_3 \cdot b_{3d}],
% \end{equation}
% % \begin{equation}
% %     \Omega_d = \begin{bmatrix}
% %         \Omega_{1d} \\\Omega_{2d} \\ \Omega_{3d}
% %     \end{bmatrix} = \begin{bmatrix}
% %         -h_{\Omega} \cdot b_{2d}\\
% %         h_{\Omega} \cdot b_{1d}\\
% %         \dot{\psi}_d e_3 \cdot b_{3d}
% %     \end{bmatrix},
% % \end{equation}
% where $h_{\Omega}$ is an intermediate vector such that $h_{\Omega} = (\dot{a}_d-(b_{3d} \cdot \dot{a}_d) b_{3d})/\norm{\ddot{p}_d+ g e_3}$ 
% % \begin{equation}
% %     % h_omega = 1/norm_t*(xd_3dot-dot(z_B,xd_3dot)*z_B);
% %     % t = [xd_2dot(1),xd_2dot(2),xd_2dot(3) + g]';
% %     % norm_t = norm(t);
% %     h_{\Omega} = \frac{\dot{a}_d-(b_{3d} \cdot \dot{a}_d) b_{3d}}{\norm{\ddot{p}_d+ g e_3}}
% % \end{equation}
% with $a_d = \ddot{p}_d$ being the desired linear acceleration and the assumption that $\norm{\ddot{p}_d+ g e_3} \neq 0$.
% The} 
and the desired moment is computed via
\begin{multline}\label{eq:torque control}
    M= -K_R e_R - K_\Omega e_\Omega + \Omega \times J \Omega  \\
     - J( \Omega^{\wedge} R^\top R_d \Omega_d-R^\top R_d \dot{\Omega}_d),
\end{multline}
where $K_R,K_{\Omega} \in \mathbb{R}^{3 \times 3}$ are user-selected positive-definite gain matrices. The derivation of $\Omega_d$ and $\dot{\Omega}_d$ are omitted in this paper for simplicity; see~\cite[Appendix F]{lee2010geometricarxiv} for details. 
% \sout{The desired angular acceleration $\dot{\Omega}_d$ is computed using the differential flatness:}
% \begin{equation}
%   \msout{ \dot{\Omega}_d^\top = [\dot{\Omega}_{1d} \ \dot{\Omega}_{2d} \ \dot{\Omega}_{3d}]=[- h_{\dot{\Omega}} \cdot b_{2d}\ h_{\dot{\Omega}} \cdot b_{1d}\        \ddot{\psi}_d e_3 \cdot b_{3d}],}
% \end{equation}

% \begin{equation}
%     \dot{\Omega}_d = \begin{bmatrix}
%         \dot{\Omega}_{1d} \\ \dot{\Omega}_{2d} \\ \dot{\Omega}_{3d}
%     \end{bmatrix} = \begin{bmatrix}
%         - h_{\dot{\Omega}} \cdot b_{2d}\\
%         h_{\dot{\Omega}} \cdot b_{1d}\\
%         \ddot{\psi}_d e_3 \cdot b_{3d}
%     \end{bmatrix},
% \end{equation}
% \sout{where $h_{\dot{\Omega}}$ is the intermediate vector}

% \begin{align}
% \msout{
%     h_{\dot{\Omega}} = & \frac{\ddot{a_d} - [(\Omega_d \times b_{3d})\cdot \ddot{p}_d + b_{3d} \cdot \ddot{a}_d ] b_{3d} +(b_{3d} \cdot \dot{a}_d)\Omega^{\wedge} b_{3d} }{\norm{\ddot{p}_d+ g e_3}} \nonumber \\
%     &-\Omega \times b_{3d} - \Omega \times(\Omega \times b_{3d}).}
% \end{align}

To summarize, the geometric controller tracks the position $p_d$ and yaw angle $\psi_d$ by setting the thrust $f$ and moment $M$ via \eqref{eq:thrust control} and \eqref{eq:torque control}, respectively. %\textcolor{red}{ZW: geometric controller also tracks velocity, correct?} {\color{blue}Sheng: I think ``tracking position'' by itself includes tracking the velocity. If position is tracked only while velocity is not, then position is not tracked.}
For the stability proof of system \eqref{equ:quadrotor dynamics} with the geometric controller \eqref{eq:thrust control} and \eqref{eq:torque control}, see~\cite[Prop.~3]{lee2010geometric}.
% \todo{We might need to replace this part by a sentence with citation.}
% The stability of system \eqref{equ:quadrotor dynamics} with the geometric controller \eqref{eq:thrust control} and \eqref{eq:torque control} has been shown in \cite{lee2010geometric}. Specifically, define the error function on $SO(3)$ by $\Psi(R,R_d) =  \text{tr}(I-R_d^\top R)/2$. If the initial condition satisfies $\Psi(R(0),R_d(0)) < 1$, then the zero equilibrium point of the error dynamics for \eqref{equ:quadrotor dynamics} is exponentially stable \cite[Prop.~2]{lee2010geometric}. Furthermore, if 
% \begin{align}
%     1 \leq & \Psi(R(0),R_d(0)) <2, \\
%     \norm{e_{\Omega}(0)}^2 < & \frac{2}{\lambda_{\max}(J)}K_R(2-  \Psi(R(0),R_d(0))),
% \end{align}
% where $\lambda_{\max}(J)$ is the largest eigenvalue of $J$, then the zero equilibrium point of the error dynamics of \eqref{equ:quadrotor dynamics} is exponentially attractive \cite[Prop.~3]{lee2010geometric}.

\section{Geometric controller with $\mathcal{L}_1$ augmentation}\label{sec:geometric+L1}
The nominal dynamics \eqref{equ:quadrotor dynamics} provide a description of a quadrotor's motion in an ideal case. However, in reality, a quadrotor's motion is affected by uncertainties and disturbances, such as propellers' aerodynamic effects, ground effect, and wind, for which establishing precise models is expensive with only marginal practical benefits. Next, we introduce the uncertainties to the state-space representation of the system.

Define state variable $x \in \mathbb{R}^{12}$ of the nominal dynamics \eqref{equ:quadrotor dynamics} by $x^\top = [p^\top \ v^\top \ \phi \ \theta \ \psi \ \Omega^\top]$ and define the partial state $z \in \mathbb{R}^6$ by $z^\top = [v^\top \  \Omega^\top]$. The state-space form of \eqref{equ:translational dynamics} and \eqref{equ:rotational dynamics} with partial state $z$ is
\begin{equation}\label{equ:nominal system with partial state}
    \dot{z}(t)=f(z(t)) +B(R(t))u_b(t),
\end{equation}
where $f(z)= \left[\begin{smallmatrix}
        ge_3 \\
        -J^{-1}\Omega \times J \Omega
    \end{smallmatrix}\right]$, $B(R)=\left[  \begin{smallmatrix}
        -m^{-1}Re_3 & 0_{3\times3}\\
        0_{3\times1} & J^{-1}
    \end{smallmatrix} \right]$,
% \begin{equation*}
%     f(z)= \begin{bmatrix}
%         ge_3 \\
%         -J^{-1}\Omega \times J \Omega
%     \end{bmatrix}, \
%     B(R)=  \begin{bmatrix}
%         -m^{-1}Re_3 & 0_{3\times3}\\
%         0_{3\times1} & J^{-1}
%     \end{bmatrix},
% \end{equation*} 
and $u_b^\top = [f \ M^\top]$ is the baseline (geometric) controller. The uncertainties enter the system \eqref{equ:quadrotor dynamics} via \eqref{equ:translational dynamics} and \eqref{equ:rotational dynamics}, which results in the uncertain dynamics:
\begin{multline}\label{equ:uncertain system with partial state}
    \dot{z}(t)=f(z(t)) + B(R(t))(u_b(t) + \sigma_m(t,x(t))) \\ + B^\bot(R(t)) \sigma_{um}(t,x(t)),
\end{multline}
where $\sigma_m \in \mathbb{R}^4$ and $\sigma_{um} \in \mathbb{R}^2$ stand for the matched and unmatched uncertainties, respectively, and $B^{\bot}(R)= \left[ \begin{smallmatrix}
        m^{-1}Re_1 &m^{-1}Re_2\\
        0_{3\times1} & 0_{3\times1}
    \end{smallmatrix}\right]$.
% \begin{equation}
%     B^{\bot}(R)= \begin{bmatrix}
%         m^{-1}Re_1 &m^{-1}Re_2\\
%         0_{3\times1} & 0_{3\times1}
%     \end{bmatrix}.
% \end{equation}
The matched uncertainty $\sigma_m$ enters the system in the same way as the control channel $u_b$ (through $B(R)$) and, hence, can be compensated by the $\mathcal{L}_1$ adaptive controller, whereas the unmatched one $\sigma_{um}$ enters the system through $B^{\bot}(R)$ whose columns are perpendicular to those of $B(R)$. A physical interpretation for uncertainties $\{\sigma_m,\sigma_{um}\}$ is the unmodelled force $F_0 \in \mathbb{R}^3$ and moment $M_0 \in \mathbb{R}^3$ applied to the COM in the body-fixed frame. In this case, the matched uncertainty $\sigma_m$ contains $F_0$'s projection onto the body-$z$ axis and $M_0$ (i.e., $\sigma_m^\top = [F_0 \cdot Re_3 \ M_0^\top]$), whereas the unmatched uncertainty~$\sigma_{um}$ contains $F_0$'s projection onto the body-$xy$ plane (i.e., $\sigma_{um}^\top = [F_0 \cdot Re_1\ F_0 \cdot R e_2]$. The uncertainties in \eqref{equ:uncertain system with partial state} also apply to the case with perturbed mass and moment of inertia (see details in \cite{pravitra2020MPPI}). Note that we do not consider input constraints in this paper. For the readers who are interested in incorporating input saturation, please refer to~\cite{li2009l1}.

Notice that both types of uncertainties, $\sigma_m(t,x(t))$ and $\sigma_{um}(t,x(t))$, are modelled as nonparametric uncertainties, i.e., they cannot be parameterized with a finite number of parameters (in the form of $\xi W(x)$ for $\xi$ being unknown parameters and $W(x)$ being known vector-valued nonlinear functions of $x$) \cite{boyd1986note}. Another important feature is that both uncertainties are time- and state-dependent, which applies to a broad class of uncertainties in practice.

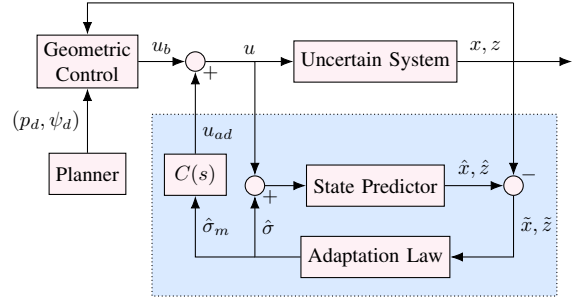
\begin{figure}[t]
    \vspace{0.5 cm}
\begin{center}
    \begin{tikzpicture}[scale=0.78, transform shape,align=center] % For altering the scale

     \filldraw[fill=lcolor!70, densely dotted] (1.1,-0.9) -- (8,-0.9) -- (8,-4) -- (1.1,-4)  -- cycle;
        % We start by placing the blocks
       \node [input, name=input] {};
        \node [block] (ccm)  {Geometric\\Control};
        \node [sum] (sum) [right=0.8cm of ccm] {};
        \node [block] (sys) [right=1.5cm of sum] {Uncertain System};
        \node [sqblock] (filt) [below=1.4cm of sum] {$C(s)$};
          \node [block] (planner) [below=1.1cm of ccm] {Planner};
        \node [block] (pred) [below=1.4cm of sys] {State Predictor};
        \node [block] (adap) [below=0.5cm of pred] {Adaptation Law};
        \node [sum] (diff) [right=1.0cm of pred] {};
        \node [sum] (sum2) [left=0.7cm of pred] {};
        \node [phantom] (p2) at (sys -| diff) {};
        \node [phantom, above=1.0cm of p2] (p3) {};
        \node [phantom, below=1.9cm of p2] (p4) {};
        \node [phantom, above=1cm of p2] (p11) {};
        \node [phantom, right=1.3cm of sum] (p6) {};
        \node [phantom, below=1.9cm of p6] (p7) {};
        \node [phantom, above=2.5cm of sys] (p8) {};
        \node [phantom] (p9) at (adap -| sum2) {};
        % \node [phantom] (p10) at (sys |- ccm) {};

        \draw[-Latex] (planner) -- node [midway,left] {$(p_d,\psi_d)$} (ccm);
        \draw[-Latex] (ccm) -- node [midway,above] {$u_{b}$} (sum);
        \draw[-Latex] (sum) -- node [midway,above] {$u$} (sys);
         \draw[-Latex] (sum) -| (sum2);
        \draw[-Latex] (filt) -- node [near start,right] {$u_{ad}$} node[pos=0.99,right] {$+$} (sum);
        \draw[-Latex] (p9) -|  node[pos=0.99,right] {$+$} node [near end,right] {$\hat{\sigma}$}(sum2);
        \draw[-Latex] (pred) -- node [midway,above] {$\hat{x},\hat{z}$} (diff);
        \draw[-Latex] (diff) |- node [near start,right] {$\tilde{x},\tilde{z}$} (adap);
        \draw[-Latex] (p2) -- node[pos=0.99,right] {$-$} (diff);
        \draw (sys) -- node [midway,above] {$x,z$} (p2);
        \node [output, right=1.0cm of p2] (output) {};
        \draw[-Latex] (p2) -- (output);
        \draw[-Latex] (sum2) -- (pred);
        \draw (p2) -- (p11);
        \draw[-Latex] (p11) -| (ccm);
        \draw (adap) -- (p9);
        % \draw (adap) -| node [near end,right] {$\hat{\sigma_m}$}(filt)
        \draw[-Latex] (p9) -| node [near end,right] {$\hat{\sigma}_m$}(filt);

    \end{tikzpicture}
    % \vspace{-0.1cm}
  \caption{The framework of geometric control with $\mathcal{L}_1$ augmentation, where the $\mathcal{L}_1$ controller is highlighted in blue. }
  \label{fig: L1 control framework}
      \vspace{-0.5cm}
 \end{center}
\end{figure}

The $\mathcal{L}_1$ adaptive controller includes a state predictor, an adaptation law, and a low-pass filter (LPF) as shown in Fig.~\ref{fig: L1 control framework}. The state predictor replicates the systems' structure, with the unknown uncertainties replaced by their estimates. If the uncertainty estimation by the adaptation law is accurate, then the prediction error should go to zero. %{\color{green}Shall we remind the readers that the state predictor is not an estimator/observer, i.e., the states are available.} % Note that the state predictor requires the full knowledge of $z$ and $R$, and it is not an estimator of the state.
The $\mathcal{L}_1$ control $u_{ad}$ is inserted to the uncertain dynamics such that
\begin{multline}\label{equ:quadrotor uncertain dynamics}
    \dot{z}(t)=f(z(t))+B(R(t))\left(u_b(t) + u_{ad}(t)+\sigma_m(t,x(t))\right) \\ +B^{\bot}(R(t))\sigma_{um}(t,x(t)),
\end{multline}
where $u_{ad} \in \mathbb{R}^4$ such that $u_{ad}^\top=[u_{f_{\mathcal{L}_1}}^\top \ u_{M_{\mathcal{L}_1}}^\top]$. 
The state predictor is 
\begin{multline}\label{equ:L1statepredictor}
    \dot{\hat{z}}(t)=f(z(t)) +B(R(t))(u_b(t) +  u_{ad}(t)+\hat{\sigma}_m(t)) \\ +B^{\bot}(R(t))\hat{\sigma}_{um}(t)+A_s\Tilde{z}(t),
\end{multline}
where $\Tilde{z}=\hat{z}-z$ is the prediction error and $A_s \in \mathbb{R}^{6 \times 6}$ is a user-selected diagonal Hurwitz matrix (which drives the prediction error $\norm{\tilde{z}}$ to 0 exponentially fast). We use the piecewise-constant adaptation law such that for $t \in [iT_s,(i+1)T_s)$
\begin{equation}\label{equ:L1adaptationlaw}
\hat{\sigma}(t)
% \begin{bmatrix}
% \hat{\sigma}_m(t)\\
% \hat{\sigma}_{um}(t)
% \end{bmatrix} 
=
\hat{\sigma}(iT_s)
% \begin{bmatrix}
% \hat{\sigma}_m(iT_s)\\
% \hat{\sigma}_{um}(iT_s)
% \end{bmatrix}
=-\bar{B}(iT_s)^{-1}\Phi^{-1}\mu(iT_s),
\end{equation}
where $\hat{\sigma}^\top = [\hat{\sigma}_m^\top \ \hat{\sigma}_{um}^\top]$, $T_s$ is the time step, $\bar{B}(iT_s)=[B(R(iT_s)) \ B^\bot(R(iT_s))]$, $\Phi=A_s^{-1}(\exp(A_sT_s)-I)$, and $\mu(iT_s)=\exp(A_sT_s)\Tilde{z}(iT_s)$ for $i \in \mathbb{N}$. 
Note that the square matrix $\bar{B}$ is invertible since it has full rank. Moreover, each element of $\bar{B}^{-1}$ has an explicit form, which enables fast computation of this matrix inverse.

The $\mathcal{L}_1$ control law only compensates for the matched uncertainty $\sigma_m$ within the bandwidth of the LPF with transfer function $C(s)$:
\begin{equation}\label{equ:L1controlinput}
u_{ad}(s)=-C(s)\hat{\sigma}_m(s),
\end{equation}
where the signals are posed in the Laplacian domain and the filter bandwidth $\omega_c$ needs to satisfy the stability conditions~\cite[Sec.~III]{lakshmanan2020safe}.

% \textcolor{blue}{$\mathcal{L}_1$ adaptive controller only compensates for matched uncertainties within the bandwidth of the LPF $C(s)$. One may increase the filter bandwidth to achieve better uncertainties attenuation. However, an LPF with high bandwidth will allow more high-frequency components to pass through to the control channel, thus damaging the robustness of the system. The bandwidth defines the tradeoff between the performance and robustness of the closed-loop system.}

The piecewise constant adaptation law can completely eliminate the state prediction error dynamics $\tilde{z}(iT_s)$ at the next sample time $(i+1)T_s$. To see how it works, consider the state prediction error via subtracting~\eqref{equ:quadrotor uncertain dynamics} from~\eqref{equ:L1statepredictor}:
\begin{equation}\label{equ:prediction error dynamics}
    \dot{\tilde{z}}(t) = A_s \tilde{z}(t) + \bar{B}(t)(\hat{\sigma}(t)-\sigma(t)),
\end{equation}
where $\sigma^\top = [{\sigma}_m^\top \ {\sigma}_{um}^\top]$. Without loss of generality, assume that the initial prediction error is nonzero, i.e., $\tilde{z}(0) \neq 0$. The closed-form solution for \eqref{equ:prediction error dynamics} is 
\begin{multline}\label{equ:closed-form solution to prediction error}
    \tilde{z}(T_s)=\exp(A_s T_s) \tilde{z}(0) + \left(\exp(A_s T_s) - I \right) A_s^{-1} \bar{B}(0) \hat{\sigma}(0) \\
    + \int_0^{T_s} \text{exp}((T_s-t)A_s) \bar{B}(0) \sigma(t) \dd t
\end{multline}
due to \eqref{equ:prediction error dynamics}'s linearity and $\bar{B}(t)$ and $\hat{\sigma}(t)$ being constants for $t \in [0,T_s)$. Setting $i=0$ in \eqref{equ:L1adaptationlaw} and plugging $\hat{\sigma}(0)$ in \eqref{equ:closed-form solution to prediction error}, the prediction error takes the form $\tilde{z}(T_s) = \int_0^{T_s} \text{exp}((T_s-t)A_s) \bar{B} \sigma(t) \dd t$, where the initial error $\tilde{z}(0)$ does not show up. Moreover, the error $\tilde{z}(T_s)$ will be eliminated in $\tilde{z}(2T_s)$ following the same logic. By setting the sampling time $T_s$ small enough (up to the hardware limit), one can keep $\norm{\tilde{z}}$ small and achieve arbitrary fast uncertainty compensation. Note that small $T_s$ may result in high adaptation gain $\Phi^{-1} \exp(A_sT_s)$ in~\eqref{equ:L1adaptationlaw}, which achieves fast estimation. Simultaneously, the high adaptation gain will introduce high-frequency components in the estimates. Therefore, we filter the uncertainty estimates to reject the high-frequency components from entering the control channel. Thus, the fast estimation is decoupled from the control channel and cannot hurt the robustness of the system.

Our design of the $\mathcal{L}_1$ controller is based on the partial state $z$ instead of the full state $x$ for the following two reasons. First, the kinematics \eqref{eq:position integral} and \eqref{eq:angle integral} are uncertainty-free. Therefore, we only consider the partial state $z$ that is directly affected by the uncertainties. 
%{\color{blue}Second, the partial state $z$ lives in the Euclidean space, where the prediction error $\tilde{z}$ is defined straightforward. When the angular kinematics \eqref{eq:angle integral} is included in the design (e.g., \cite{kotaru2020geometric}), care has to be taken to the (angular) error which is defined in $SO(3)$. Such a design introduces unnecessary complexities since the kinematics are uncertainty-free. }
Next, if we use the full state $x$, then the matrix $\bar{B} = [B(R) \ B^\bot(R)] \in \mathbb{R}^{12 \times 6}$ is not square, which results in the infeasibility of \eqref{equ:L1adaptationlaw} since $\bar{B}^{-1}$ is not well-defined. For complete proofs of stability and performance of $\mathcal L_1$ adaptive controller refer to~\cite{hovakimyan2010L1}.

\section{Experimental results}\label{sec:experiments}
% \pz{Some figures showing the desired and actual trajectories with/without L1 should be included. The RMSE plot error is not enough to give a {\it } visual impression. One can pick one to two uncertainty cases due to space limit}

We demonstrate the proposed $\mathcal{L}_1$ augmentation on a Mambo quadrotor (displayed in Fig.~\ref{fig:mambo with frames}) in experiments. 
We design our controller in Simulink and generate C code for the quadrotor's onboard execution~\cite{MamboSimulink}. 
All the computations including state estimation, baseline and $\mathcal{L}_1 $ control calculation, and data logging run onboard at 200~Hz.
% The quadrotor runs at 200~Hz, which includes estimating the vehicle states using a complementary filter~\cite{mahony2008nonlinear} and Kalman filter; computing the baseline (geometric) and $\mathcal{L}_1 $ controls; and logging the data. 
For the LPF, we use a first-order LPF for the thrust channel and two cascaded first-order LPFs for the moment channel. 
The position of the quadrotor is provided from Vicon cameras at 120~Hz. We set $\Omega_d = \dot{\Omega}_d$ = 0 for simplicity in implementation. A list of parameters in the experiments is shown in Table~\ref{tb: parameters in experiments}. %The coefficients $k_F$, $k_M$ for motor mixing \eqref{eq:motor mapping to thrust and moment} is provided by in the Simulink model~\cite{MamboSimulink}.

\setlength{\tabcolsep}{5pt} % Default value: 6pt
\renewcommand{\arraystretch}{1} % Default value: 1
  \captionsetup{%size=footnotesize,
	%justification=centering, %% not needed
	skip=5pt, position = bottom}
\begin{table}[!b]
	\centering
	\small
	
	\vspace{-0.5cm}
	
	%\captionsetup{font=small}
	\caption{Parameters in experiments.}
	\begin{tabular}{llcll}
		\toprule[1pt]
		 param. & value & & param. & value \\
		\cmidrule{1-2} \cmidrule{4-5} 
        $m$  & 0.075 kg & & $J$ & $10^{-5}\diag{5.8 \ 7.2 \ 10}$ kgm$^2$ \\
        g & 9.81 m/s$^2$ & & $K_p$ & $10^{-1}\diag{6.2 \ 4.5 \ 6.9}$\\
        $T_s$ & 0.005 s & & $K_v$ & $10^{-1}\diag{1 \ 1 \ 1.5}$ \\
        $\omega_c$ ($f$) & 8  rad/s && $K_R$ & $10^{-2}\diag{1.5 \ 1.5 \ 0.2} $ \\
        $\omega_c$ ($M$) & 4, 6 rad/s & & $K_{\Omega}$ & $10^{-3}\diag{2.2 \ 2.2 \ 0.7} $  \\% filter bandwidth 
        & && $A_s$ & $-\diag{5 \ 5 \ 5 \ 10 \ 10 \ 10}$ \\
		\bottomrule[1pt]
	\end{tabular}\label{tb: parameters in experiments}
% 	\vspace{-0.3cm}
\end{table}
\normalsize

\subsection{Uncertainty estimation and compensation}

% \textcolor{red}{Adi: Is artificial uncertainty being injected? In this case the results should be sold as `Hardware in Loop' simulation.} \textcolor{blue}{ZW:since we are doing actual flight test, this is not considered as Hardware in the Loop simulation, I confirmed this with Kasey and Kevin.} \textcolor{red}{Adi: Great!}

\begin{figure}

\vspace{0.2cm}

     \begin{subfigure}[b]{\columnwidth}
         \centering
         \includegraphics[width=\columnwidth]{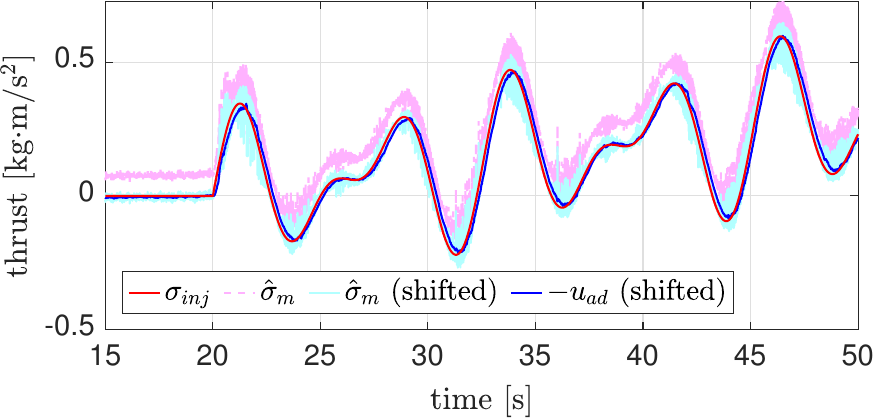}
         \caption{Injected uncertainty $\sigma_{inj}$ on thrust.}
        \label{fig:hover thrust}
     \end{subfigure}
     \\
      \begin{subfigure}[b]{\columnwidth}
         \centering
         \includegraphics[width=\columnwidth]{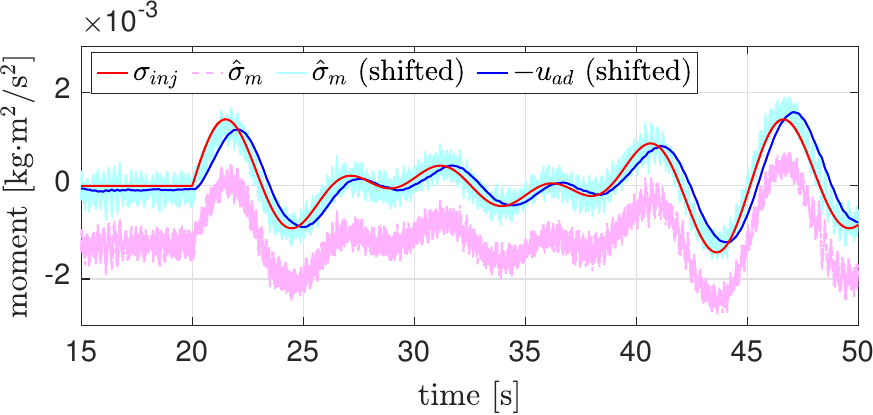}
         \caption{Injected uncertainty $\sigma_{inj}$ on pitch moment.}
        \label{fig:hover moment}
     \end{subfigure}
     \caption{$\mathcal{L}_1$ controller's uncertainty estimate $\hat{\sigma}_m$ and compensation $-u_{ad}$ versus the injected uncertainty $\sigma_{inj}$. We shift $\hat{\sigma}_m$ and $-u_{ad}$ vertically to show their convergence to $\sigma_{inj}$.}
     
     \vspace{-0.5cm}
\end{figure}

We illustrate the $\mathcal{L}_1$ augmentation's capability of handling uncertainty by injecting signals to the control channel. The injected signals, denoted by $\sigma_{inj}$, serve as artificial uncertainties such that the total (matched) uncertainty is the sum of existing uncertainties in the system and the injected $\sigma_{inj}$. 
% Both the baseline and $\mathcal{L}_1$ controller are agnostic to $\sigma_m$ \textcolor{red}{(What does `agnostic' imply here?)}. 
The uncertainty $\sigma_m$ should be estimated by the adaptation law by $\hat{\sigma}_m$ and then compensated by the $\mathcal{L}_1$ controller. For illustration purposes, we set the first element of $\sigma_{\rm inj}(t)$ to be $0.2\sin(t-20) + 0.01(t-20) + 0.15\sin(1.5(t-20))$ for $t \in [20,50]$, and the other elements of $\sigma_{inj}(t)$ are set to 0 for the entire time. In other words, the injected uncertainty only applies to the thrust from 20 to 50 s. We set the quadrotor to hover at 1~m altitude so that the uncertainty estimate's convergence to the injected uncertainty can be clearly observed. 
The result is shown in Fig.~\ref{fig:hover thrust}.
The uncertainty estimate $\hat{\sigma}_m$ quickly converges to the injected signal $\sigma_{inj}$, and $u_{ad}$ compensates for the injected uncertainty.
Figure~\ref{fig:hover moment} shows similar uncertainty estimation and compensation results for the case when the injected uncertainty only applies to the third element of $\sigma_m$ (pitch moment), where the injected signal is $10^{-3}(2\sin(t-20) + 0.2(t-20) + \sin(0.75(t-20))$ for $t \in [20,50]$. Note that the injected signals in this subsection, and later in Table~\ref{tab:injected uncertainties in trajectory tracking}, have a significant impact to the quadrotor's linear acceleration $\dot{v}$ (angular acceleration~$\dot{\Omega}$) since $\sigma_{\rm inj}$ is scaled by $m^{-1}$ ($J^{-1}$).

\subsection{Tracking performance}
\begin{figure}
    \centering
    \includegraphics[width=\columnwidth]{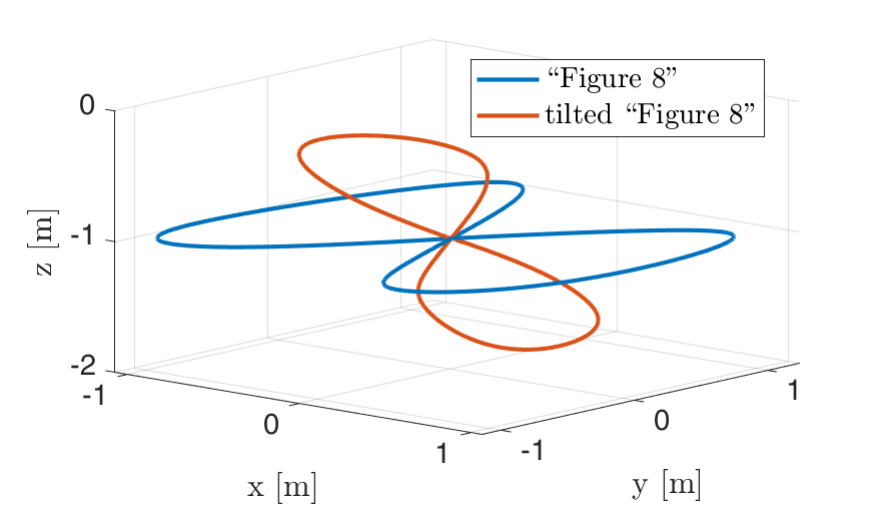}
    \caption{Desired trajectories in Experiments 1 and 2.}
    \label{fig:figure8 and titled figure8}
    
    \vspace{-0.5cm}
\end{figure}

We compare the performance of the $\mathcal{L}_1$ augmentation to that of the baseline controller only. We deploy the controllers to track a ``Figure 8'' and a tilted ``Figure 8''-shaped trajectory (shown in Fig.~\ref{fig:figure8 and titled figure8}) in Experiments 1 and 2, respectively. In Experiment~1, we consider the following six cases: 
% 1) No artificial uncertainty; 2) Time-dependent uncertainty (see Table~\ref{tab:injected uncertainties in trajectory tracking}); 3) State-dependent uncertainty (see Table~\ref{tab:injected uncertainties in trajectory tracking}); 4) Unknown input gain $\beta(t) = 1+0.4\sin(t-5)$ on the pitch moment for $t\in [5,22]$; 5) Chipped propeller: propeller on motor 1 is cut off by 0.9 cm (14\% of a propeller); 6) Poorly tuned baseline controller. 
\begin{enumerate}[leftmargin=*]
    \item no artificial uncertainty;
    \item time-dependent uncertainty (see Table~\ref{tab:injected uncertainties in trajectory tracking}); % $\sigma_{inj}^\top(t) = 10^{-3}[0 \ \sin(0.75(t-5)) \ 0.8\sin(t-5)+0.8\sin(0.5(t-5)) \ 0]$ for $t\in [5,22]$.
    \item state-dependent uncertainty (see Table~\ref{tab:injected uncertainties in trajectory tracking}); %$\sigma_{\rm s}^\top(t) = 10^{-3}[0 \ \sin(0.75(t-5)) \ 0.5\sin(t-5)p_1^2(t) \ 0]$ for $t\in [5,22]$.
    \item unknown input gain $\beta(t) = 1+0.4\sin(t-5)$ on the pitch moment for $t\in [5,22]$;
    \item chipped propeller: propeller on motor 1 is cut off by 0.9~cm (14\% of a propeller); 
    \item poorly tuned baseline controller. 
\end{enumerate}
The time- and state-dependent artificial uncertainties in Cases~2 and 3 are injected into the control channel. The unknown input gain $\beta$ applies to the input matrix $B(R)$ such that the actual input to the system in \eqref{equ:quadrotor uncertain dynamics} is multiplied by $\beta B(R)$, which associates with the scenario where the input matrix has time-varying model uncertainties. 
% despite pooly tuned xxx, l1 can still achieve the same level of performance
% to illustrate how effective l1 is, we deliberately
Tuning a nonlinear controller is a challenging task, which requires significant model knowledge and can be time consuming~\cite{berkenkamp2016safe}. Case 6 is deliberately included to demonstrate that the $\mathcal{L}_1$ augmentation can achieve the same level of performance despite a poorly tuned baseline controller. 
% Case 6 is considered because a poorly tuned baseline controller may be implemented in practice due to difficulties in tuning nonlinear controllers (e.g., the geometric controller) and limited knowledge of a nonlinear system. 
In our case, the poorly tuned baseline controller uses smaller gains $K_p$ and $K_R$ than those in Table~\ref{tb: parameters in experiments}, which results in significantly larger steady-state error and slower response to commands. Note that Cases 4 and 6 do not apply to the uncertainty model \eqref{equ:uncertain system with partial state}: we use them to empirically demonstrate the performance of $\mathcal{L}_1$ augmentation.

\begin{figure}[!t]
    \begin{subfigure}[t]{\columnwidth}
       \centering
         \includegraphics[width=\columnwidth]{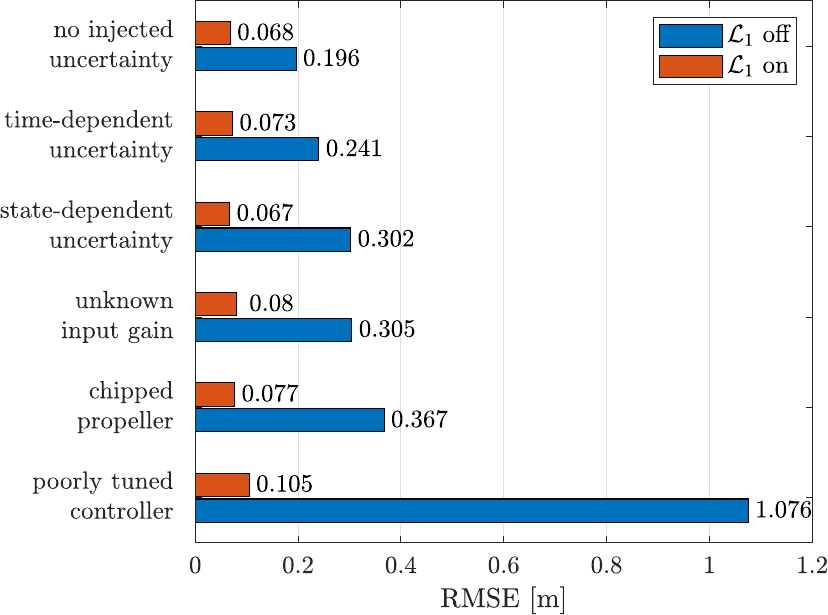}
         \caption{Experiment 1 (``Figure 8'' trajectory).}
        \label{fig: bar plot figure 8}
    \end{subfigure}\\
    \begin{subfigure}[t]{\columnwidth}
       \centering
         \includegraphics[width=\columnwidth]{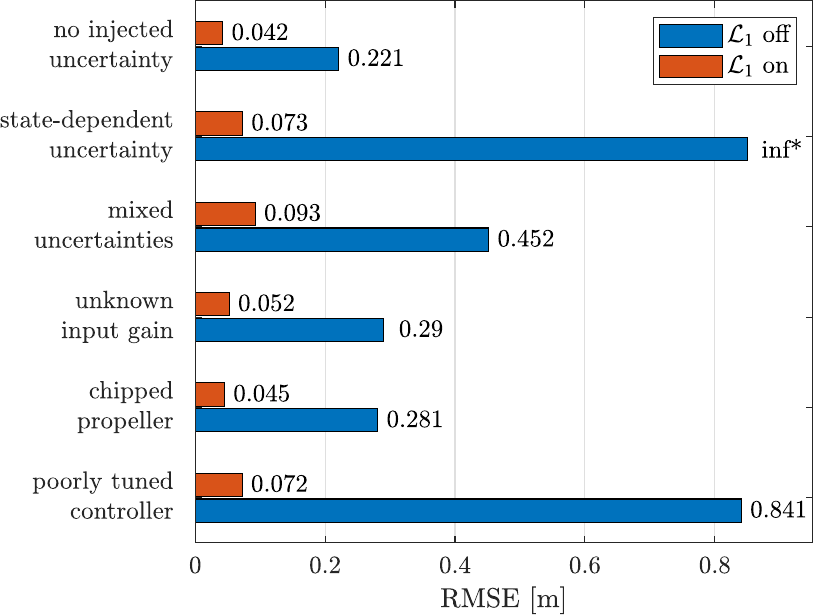}
         \caption{Experiment 2 (tilted ``Figure 8'' trajectory). The ``inf*'' mark refers to the failed trial using the baseline controller alone under state-dependent uncertainty.}
        \label{fig: bar plot tilted figure 8}
    \end{subfigure}
    \caption{Comparisons of tracking error under various types of uncertainties between the baseline controller alone ($\mathcal{L}_1$ off) and the baseline controller with $\mathcal{L}_1$ augmentation ($\mathcal{L}_1$ on).}
    \label{fig: bar plots}
    
    \vspace{-0.5cm}
\end{figure}

\begin{figure}[!t]
% \vspace{-0.5cm}

    \centering
    \includegraphics[width=\columnwidth]{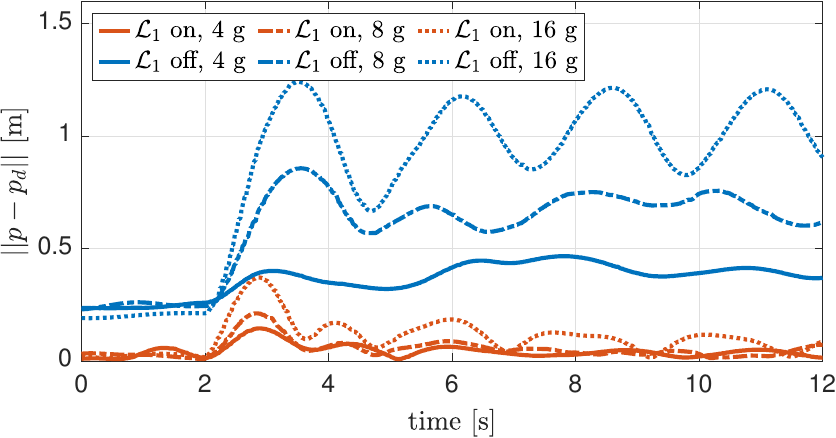}
    \caption{Comparison of tracking errors for hovering in the cases with and without $\mathcal{L}_1$ augmentation under added slung weights. The weights are added at 2 s. The 4/8/16~g weights correspond to 5.3/10.7/21.3\% of the quadrotor's weight.}
    \label{fig: adding weights}
    
    \vspace{-0.5cm}
\end{figure}

% \begin{figure}[!t]
% % \vspace{-0.5cm}

%     \centering
%     \includegraphics[width=\columnwidth]{Figure8HighGain_L1OnOff_GoodOrChippedProp_distance.eps}
%     \caption{Comparison of trajectory tracking error for the cases with and without $\mathcal{L}_1$ augmentation, and without and with a chipped propeller.}
%     \label{fig: l1 compared with high-gain baseline}
    
%     \vspace{-0.5cm}
% \end{figure}

\setlength{\tabcolsep}{5pt} % Default value: 6pt
\renewcommand{\arraystretch}{1} % Default value: 1
  \captionsetup{%size=footnotesize,
	%justification=centering, %% not needed
	skip=5pt, position = bottom}
\begin{table*}[!t]
	\centering
	\small
	
% 	\vspace{-0.3cm}
	
	%\captionsetup{font=small}
	\caption{List of injected uncertainties in Experiments 1 and 2 for time $t \in [5, 22]$.}
	\begin{tabular}{ccccc}
		\toprule[1pt]
		 Exp. & Case & $\sigma_m$ (thrust) &$\sigma_m$ (roll moment) & $\sigma_m$ (pitch moment)\\
		 \midrule
		 1 & 2 & 0 & $0.001\sin(0.75(t-5))$ & $ 0.0008\sin(t-5)+0.0008\sin(0.5(t-5))$  \\
		 1 & 3 & 0 & $0.001\sin(0.75(t-5))$ & $ 0.0005\sin(t-5)p_1^2(t)$ \\
		 2 & 2 & 0 & $0.001\sin(0.75(t-5))$ & $0.01\sin(t-5)p_1^2(t)$ \\
		 2 & 3 & $0.2\sin(0.5(t-5)) +0.15\sin(0.75(t-5))$ & $0.001\sin(0.75(t-5))$ & $0.0008(\sin(t-5)+\sin(0.5(t-5)))$\\
		\bottomrule[1pt]
	\end{tabular}
	\label{tab:injected uncertainties in trajectory tracking}
	
	\vspace{-0.5cm}
\end{table*}
\normalsize
The root-mean-square error (RMSE) for trajectory tracking is shown in Fig.~\ref{fig: bar plot figure 8}. The $\mathcal{L}_1$ augmentation can maintain a consistent and much smaller RMSE tracking error compared to that of the baseline controller alone under various types of uncertainties. 
% We compare the tracking error for Cases 1 and 5 in Fig.~\ref{fig: l1 compared with high-gain baseline}. The comparison demonstrates that the $\mathcal{L}_1$ augmentation can compensate for the thrust loss due to the chipped propeller and also maintain an almost same level of tracking error compared with the case using full propellers.

% In Experiment~2, we consider the following six cases:
% \begin{enumerate}[leftmargin=*]
%     \item no injected uncertainty;
%     \item state-dependent uncertainty (see Table~\ref{tab:injected uncertainties in trajectory tracking}); %$\sigma_{\rm s}^\top(t) = 10^{-2}[0 \ 0.1\sin(0.75(t-5)) \ \sin(t-5)p_1^2(t) \ 0]$ for $t\in [5,22]$.
%     \item mixed uncertainties (see Table~\ref{tab:injected uncertainties in trajectory tracking}) with unknown input gain $\beta(t) = 1+0.2\sin(t-5)$ on the thrust for $t\in [5,22]$;% $\sigma_{\rm mix}^\top(t) = 10^{-3}[200\sin(0.5(t-5)) +150\sin(0.75(t-5)) \  \sin(0.75(t-5)) \ 0.8\sin(t-5)+0.8\sin(0.5(t-5)) \ 0]$ with unknown input gain $1+0.2\sin(t-5)$ for $t\in [5,22]$.
% % additive uncertainty on thrust: y=[];
% %                         additive uncertainty on moments: y=[0.001*sin(0.75*(u-5));0.0008*sin(u-5)+0.0008*sin(0.5*(u-5));0];
% %                         unknown input gain:     y=(1+0.2*sin(t-5))*u;  where u is the nomianl thrust
%     \item unknown input gain $\beta(t) = 1+0.2\sin(t-5)$ on the thrust for $t\in [5,22]$;
%     \item chipped propeller: propeller on motor 1 is cut off by 0.9~cm (14\% of a propeller); 
%     \item poorly tuned baseline controller.
% \end{enumerate}
In Experiment~2, we consider the same cases as in Experiment~1 except:
\begin{enumerate}[leftmargin=*]
    \setcounter{enumi}{1}
    % \item state-dependent uncertainty (see Table~\ref{tab:injected uncertainties in trajectory tracking}); %$\sigma_{\rm s}^\top(t) = 10^{-2}[0 \ 0.1\sin(0.75(t-5)) \ \sin(t-5)p_1^2(t) \ 0]$ for $t\in [5,22]$.
    \item mixed uncertainties (see Table~\ref{tab:injected uncertainties in trajectory tracking}) with unknown input gain $\beta(t) = 1+0.2\sin(t-5)$ on the thrust for $t\in [5,22]$;% $\sigma_{\rm mix}^\top(t) = 10^{-3}[200\sin(0.5(t-5)) +150\sin(0.75(t-5)) \  \sin(0.75(t-5)) \ 0.8\sin(t-5)+0.8\sin(0.5(t-5)) \ 0]$ with unknown input gain $1+0.2\sin(t-5)$ for $t\in [5,22]$.
% additive uncertainty on thrust: y=[];
%                         additive uncertainty on moments: y=[0.001*sin(0.75*(u-5));0.0008*sin(u-5)+0.0008*sin(0.5*(u-5));0];
%                         unknown input gain:     y=(1+0.2*sin(t-5))*u;  where u is the nomianl thrust
    \item unknown input gain $\beta(t) = 1+0.2\sin(t-5)$ on the thrust for $t\in [5,22]$;
\end{enumerate}
The RMSE for trajectory tracking is shown in Fig~\ref{fig: bar plot tilted figure 8}. The $\mathcal{L}_1$ augmentation can maintain a relatively small and consistent RMSE on trajectory tracking, whereas the performance of the baseline controller alone suffers from uncertainties. Note that for the case with state-dependent uncertainty, the baseline controller alone crashed the quadrotor in all trials, whereas the case with the $\mathcal{L}_1$ augmentation can stabilize the quadrotor and maintain a good tracking performance.

We add different slung weights to the quadrotor during hover and compare the position-holding performance between the $\mathcal{L}_1$ augmentation and the baseline controller alone. Figure~\ref{fig: adding weights} shows the tracking errors. The $\mathcal{L}_1$ augmentation quickly adapts to the added weights and corrects the position errors, whereas the baseline controller can barely hold its position.

It worth noting that the $\mathcal{L}_1$ controller uses the same set of parameters in all cases and in all experiments (the baseline controller uses the parameters in Table~\ref{tb: parameters in experiments} for Cases 1--5). The relatively small and consistent tracking errors with the $\mathcal{L}_1$ augmentation shown in Fig.~\ref{fig: bar plots} demonstrate the superior performance and ease of use of the $\mathcal{L}_1$ controller under various types of uncertainties and for different trajectories without any redesign or retuning. Furthermore, when tracking trajectories, our approach handles the uncertainties by design, which does not require data collection and (neural network) training that are
typically used in ML-based approaches, e.g., \cite{torrente2021data,shi2020neural}.

%\section*{ACKNOWLEDGMENT}

% \textcolor{green}{Adi: Make sure the references are added correctly the first time to prevent overload at a later time. E.g., [1] needs capitalized UAV and SE.}
% {\color{blue}Sheng: Sure. We'll check it before submission.}

\section{Conclusion}\label{sec:conclusions}

In this paper, we develop, test, and validate an $\mathcal{L}_1$ adaptive controller design for a quadrotor to track a prescribed trajectory. An $\mathcal{L}_1$ augmentation of a baseline geometric controller is formulated, where the latter stabilizes the quadrotor on $SE(3)$ and the former compensates for the uncertainties and disturbances acting on the system. The $\mathcal{L}_1$ controller is based on the quadrotor's partial state that is directly affected by the uncertainties, which allows using a piecewise-constant adaptation law for uncertainty estimation. Experimental results on a Mambo quadrotor show that the uncertainty estimate quickly converges to the uncertainty, and the matched uncertainty is completely compensated for upon estimation. The $\mathcal{L}_1$ augmentation also demonstrates consistent and smaller (five times smaller on average) trajectory tracking errors compared with the baseline controller alone for different target trajectories and under various types of uncertainties.

Future work includes modifying the current design to  attenuate the unmatched uncertainties \cite{zhao2021RALPV} % {\bf i thought our existing theory supports the stability guarantees; why do you propose it as future work? Didn't you apply the L1 controller from Section III of the book for unmatched?}. 
% establishing stability guarantees for the proposed controller
and applying learning techniques (e.g., \cite{gahlawat2021contraction}) to better characterize the uncertainties using data. Learned information could be incorporated in the proposed control framework, which alleviates workloads for $\mathcal{L}_1$ adaptive controller and achieves better performance without sacrificing the robustness of the system. 

%\textcolor{red}{Adi: In place of the last para, I'd suggest: Future work would involve establishing the theoretical guarantees for the proposed controller. In our previous work in [Arun CDC] we have shown how the Riemannian energy can be used as an incremental LF for the L1 augmentation. Similarly, we will use the LF provided by the geometric control to establish the theoretical guarantees. Additionally, the proposed control architecture is amenable for the incorporation of data-driven learning methods. We have previously used Bayesian learning with L1 in [L4dc 20, 21]}. 

\bibliographystyle{IEEEtran}
\bibliography{ref}

\begin{thebibliography}{10}
\providecommand{\url}[1]{#1}
\csname url@samestyle\endcsname
\providecommand{\newblock}{\relax}
\providecommand{\bibinfo}[2]{#2}
\providecommand{\BIBentrySTDinterwordspacing}{\spaceskip=0pt\relax}
\providecommand{\BIBentryALTinterwordstretchfactor}{4}
\providecommand{\BIBentryALTinterwordspacing}{\spaceskip=\fontdimen2\font plus
\BIBentryALTinterwordstretchfactor\fontdimen3\font minus
  \fontdimen4\font\relax}
\providecommand{\BIBforeignlanguage}[2]{{%
\expandafter\ifx\csname l@#1\endcsname\relax
\typeout{** WARNING: IEEEtran.bst: No hyphenation pattern has been}%
\typeout{** loaded for the language `#1'. Using the pattern for}%
\typeout{** the default language instead.}%
\else
\language=\csname l@#1\endcsname
\fi
#2}}
\providecommand{\BIBdecl}{\relax}
\BIBdecl

\bibitem{yang2009implementation}
H.~C. Yang, R.~AbouSleiman, B.~Sababha, E.~Gjioni, D.~Korff, and O.~Rawashdeh,
  ``Implementation of an autonomous surveillance quadrotor system,'' in
  \emph{Proceedings of AIAA Infotech@Aerospace Conference}, Seattle, WA, USA,
  2009, p. 2047.

\bibitem{tian2020search}
Y.~Tian, K.~Liu, K.~Ok, L.~Tran, D.~Allen, N.~Roy, and J.~P. How, ``Search and
  rescue under the forest canopy using multiple {UAV}s,'' \emph{The
  International Journal of Robotics Research}, vol.~39, no. 10-11, pp.
  1201--1221, 2020.

\bibitem{valente2013near}
J.~Valente, D.~Sanz, J.~Del~Cerro, A.~Barrientos, and M.~{\'A}. de~Frutos,
  ``Near-optimal coverage trajectories for image mosaicing using a mini
  quad-rotor over irregular-shaped fields,'' \emph{Precision Agriculture},
  vol.~14, no.~1, pp. 115--132, 2013.

\bibitem{mellinger2013cooperative}
D.~Mellinger, M.~Shomin, N.~Michael, and V.~Kumar, ``Cooperative grasping and
  transport using multiple quadrotors,'' in \emph{Distributed Autonomous
  Robotic Systems}.\hskip 1em plus 0.5em minus 0.4em\relax Berlin, Germany:
  Springer, 2013, pp. 545--558.

\bibitem{mellinger2012trajectory}
D.~Mellinger, N.~Michael, and V.~Kumar, ``Trajectory generation and control for
  precise aggressive maneuvers with quadrotors,'' \emph{The International
  Journal of Robotics Research}, vol.~31, no.~5, pp. 664--674, 2012.

\bibitem{tayebi2006attitude}
A.~Tayebi and S.~McGilvray, ``Attitude stabilization of a {VTOL} quadrotor
  aircraft,'' \emph{IEEE Transactions on Control Systems Technology}, vol.~14,
  no.~3, pp. 562--571, 2006.

\bibitem{bhat2000topological}
S.~P. Bhat and D.~S. Bernstein, ``A topological obstruction to continuous
  global stabilization of rotational motion and the unwinding phenomenon,''
  \emph{Systems \& Control Letters}, vol.~39, no.~1, pp. 63--70, 2000.

\bibitem{mayhew2011quaternion}
C.~G. Mayhew, R.~G. Sanfelice, and A.~R. Teel, ``Quaternion-based hybrid
  control for robust global attitude tracking,'' \emph{IEEE Transactions on
  Automatic Control}, vol.~56, no.~11, pp. 2555--2566, 2011.

\bibitem{mellinger2011minimum}
D.~Mellinger and V.~Kumar, ``Minimum snap trajectory generation and control for
  quadrotors,'' in \emph{Proceedings of IEEE International Conference on
  Robotics and Automation}, Shanghai, China, 2011, pp. 2520--2525.

\bibitem{lee2010geometric}
T.~Lee, M.~Leok, and N.~H. McClamroch, ``Geometric tracking control of a
  quadrotor {UAV} on {SE}(3),'' in \emph{Proceedings of the 49th IEEE
  Conference on Decision and Control}, Atlanta, GA, USA, 2010, pp. 5420--5425.

\bibitem{emran2018review}
B.~J. Emran and H.~Najjaran, ``A review of quadrotor: {An} underactuated
  mechanical system,'' \emph{Annual Reviews in Control}, vol.~46, pp. 165--180,
  2018.

\bibitem{dydek2012adaptive}
Z.~T. Dydek, A.~M. Annaswamy, and E.~Lavretsky, ``Adaptive control of quadrotor
  {UAV}s: A design trade study with flight evaluations,'' \emph{IEEE
  Transactions on Control Systems Technology}, vol.~21, no.~4, pp. 1400--1406,
  2012.

\bibitem{selfridge2014multivariable}
J.~M. Selfridge and G.~Tao, ``A multivariable adaptive controller for a
  quadrotor with guaranteed matching conditions,'' \emph{Systems Science \&
  Control Engineering}, vol.~2, no.~1, pp. 24--33, 2014.

\bibitem{antonelli2017adaptive}
G.~Antonelli, E.~Cataldi, F.~Arrichiello, P.~R. Giordano, S.~Chiaverini, and
  A.~Franchi, ``Adaptive trajectory tracking for quadrotor {MAV}s in presence
  of parameter uncertainties and external disturbances,'' \emph{IEEE
  Transactions on Control Systems Technology}, vol.~26, no.~1, pp. 248--254,
  2017.

\bibitem{cabecinhas2014nonlinear}
D.~Cabecinhas, R.~Cunha, and C.~Silvestre, ``A nonlinear quadrotor trajectory
  tracking controller with disturbance rejection,'' \emph{Control Engineering
  Practice}, vol.~26, pp. 1--10, 2014.

\bibitem{bouabdallah2005backstepping}
S.~Bouabdallah and R.~Siegwart, ``Backstepping and sliding-mode techniques
  applied to an indoor micro quadrotor,'' in \emph{Proceedings of IEEE
  International Conference on Robotics and Automation}.\hskip 1em plus 0.5em
  minus 0.4em\relax Barcelona, Spain: IEEE, 2005, pp. 2247--2252.

\bibitem{labbadi2019robust}
M.~Labbadi and M.~Cherkaoui, ``Robust adaptive backstepping fast terminal
  sliding mode controller for uncertain quadrotor {UAV},'' \emph{Aerospace
  Science and Technology}, vol.~93, p. 105306, 2019.

\bibitem{ahmed2020disturbance}
N.~Ahmed, M.~Chen, and S.~Shao, ``Disturbance observer based tracking control
  of quadrotor with high-order disturbances,'' \emph{IEEE Access}, vol.~8, pp.
  8300--8313, 2020.

\bibitem{castillo2019disturbance}
A.~Castillo, R.~Sanz, P.~Garcia, W.~Qiu, H.~Wang, and C.~Xu, ``Disturbance
  observer-based quadrotor attitude tracking control for aggressive
  maneuvers,'' \emph{Control Engineering Practice}, vol.~82, pp. 14--23, 2019.

\bibitem{goodarzi2015geometric}
F.~A. Goodarzi, D.~Lee, and T.~Lee, ``Geometric adaptive tracking control of a
  quadrotor unmanned aerial vehicle on {SE}(3) for agile maneuvers,''
  \emph{Journal of Dynamic Systems, Measurement, and Control}, vol. 137, no.~9,
  p. 091007, 2015.

\bibitem{shi2020neural}
G.~Shi, W.~H{\"o}nig, Y.~Yue, and S.-J. Chung, ``Neural-swarm: Decentralized
  close-proximity multirotor control using learned interactions,'' in
  \emph{Proceedings of IEEE International Conference on Robotics and
  Automation}.\hskip 1em plus 0.5em minus 0.4em\relax Virtual Conference: IEEE,
  2020, pp. 3241--3247.

\bibitem{bouffard2012learning}
P.~Bouffard, A.~Aswani, and C.~Tomlin, ``Learning-based model predictive
  control on a quadrotor: Onboard implementation and experimental results,'' in
  \emph{Proceedings of IEEE International Conference on Robotics and
  Automation}.\hskip 1em plus 0.5em minus 0.4em\relax Saint Paul, Minnesota,
  USA: IEEE, 2012, pp. 279--284.

\bibitem{torrente2021data}
G.~Torrente, E.~Kaufmann, P.~F{\"o}hn, and D.~Scaramuzza, ``Data-driven {MPC}
  for quadrotors,'' \emph{IEEE Robotics and Automation Letters}, vol.~6, no.~2,
  pp. 3769--3776, 2021.

\bibitem{joshi2020asynchronous}
G.~Joshi, J.~Virdi, and G.~Chowdhary, ``Asynchronous deep model reference
  adaptive control,'' \emph{arXiv:2011.02920}, 2020, accepted for publication
  in Conference on Robots Learning 2020.

\bibitem{hovakimyan2010L1}
N.~Hovakimyan and C.~Cao, \emph{$\mathcal{L}_1$ Adaptive Control Theory:
  {G}uaranteed Robustness with Fast Adaptation}.\hskip 1em plus 0.5em minus
  0.4em\relax Philadelphia, PA, USA: SIAM, 2010.

\bibitem{gregory2009l1}
I.~Gregory, C.~Cao, E.~Xargay, N.~Hovakimyan, and X.~Zou, ``$\mathcal{L}_1$
  adaptive control design for {NASA AirSTAR} flight test vehicle,'' in
  \emph{Proceedings of AIAA Guidance, Navigation, and Control Conference},
  Chicago, IL, USA, 2009, p. 5738.

\bibitem{gregory2010flight}
I.~Gregory, E.~Xargay, C.~Cao, and N.~Hovakimyan, ``{Flight test of an
  $\mathcal{L}_1$ adaptive controller on the NASA AirSTAR flight test
  vehicle},'' in \emph{Proceedings of AIAA Guidance, Navigation, and Control
  Conference}, Toronto, Ontario, Canada, 2010, p. 8015.

\bibitem{ackerman2016l1}
K.~Ackerman, E.~Xargay, R.~Choe, N.~Hovakimyan, M.~C. Cotting, R.~B. Jeffrey,
  M.~P. Blackstun, T.~P. Fulkerson, T.~R. Lau, and S.~S. Stephens,
  ``$\mathcal{L}_1$ stability augmentation system for {Calspan's}
  variable-stability {Learjet},'' in \emph{Proceedings of AIAA Guidance,
  Navigation, and Control Conference}, San Diego, CA, USA, 2016, p. 0631.

\bibitem{ackerman2017evaluation}
K.~A. Ackerman, E.~Xargay, R.~Choe, N.~Hovakimyan, M.~C. Cotting, R.~B.
  Jeffrey, M.~P. Blackstun, T.~P. Fulkerson, T.~R. Lau, and S.~S. Stephens,
  ``Evaluation of an $\mathcal{L}_1$ adaptive flight control law on
  {Calspan’s Variable-Stability Learjet},'' \emph{AIAA Journal of Guidance,
  Control, and Dynamics}, vol.~40, no.~4, pp. 1051--1060, 2017.

\bibitem{kaminer2010path}
I.~Kaminer, A.~Pascoal, E.~Xargay, N.~Hovakimyan, C.~Cao, and V.~Dobrokhodov,
  ``Path following for small unmanned aerial vehicles using $\mathcal{L}_1$
  adaptive augmentation of commercial autopilots,'' \emph{AIAA Journal of
  Guidance, Control, and Dynamics}, vol.~33, no.~2, pp. 550--564, 2010.

\bibitem{kaminer2012time}
I.~Kaminer, E.~Xargay, V.~Cichella, N.~Hovakimyan, A.~M. Pascoal, A.~P. Aguiar,
  V.~Dobrokhodov, and R.~Ghabcheloo, ``Time-critical cooperative path following
  of multiple {UAVs}: {Case} studies,'' in \emph{Itzhack Y. Bar-Itzhack
  Memorial Symposium on Estimation, Navigation, and Spacecraft Control}.\hskip
  1em plus 0.5em minus 0.4em\relax Berlin, Germany: Springer, 2012, pp.
  209--233.

\bibitem{jafarnejadsani2017optimized}
H.~Jafarnejadsani, D.~Sun, H.~Lee, and N.~Hovakimyan, ``Optimized
  $\mathcal{L}_1$ adaptive controller for trajectory tracking of an indoor
  quadrotor,'' \emph{Journal of Guidance, Control, and Dynamics}, vol.~40,
  no.~6, pp. 1415--1427, 2017.

\bibitem{zuo2014augmented}
Z.~Zuo and P.~Ru, ``Augmented $\mathcal{L}_1$ adaptive tracking control of
  quad-rotor unmanned aircrafts,'' \emph{IEEE Transactions on Aerospace and
  Electronic Systems}, vol.~50, no.~4, pp. 3090--3101, 2014.

\bibitem{huynh20141}
M.~Q. Huynh, W.~Zhao, and L.~Xie, ``$\mathcal{L}_1$ adaptive control for
  quadcopter: Design and implementation,'' in \emph{Proceedings of the 13th
  International Conference on Control Automation Robotics \& Vision}.\hskip 1em
  plus 0.5em minus 0.4em\relax Singapore: IEEE, 2014, pp. 1496--1501.

\bibitem{kotaru2020geometric}
P.~Kotaru, R.~Edmonson, and K.~Sreenath, ``Geometric $\mathcal{L}_1$ adaptive
  attitude control for a quadrotor unmanned aerial vehicle,'' \emph{Journal of
  Dynamic Systems, Measurement, and Control}, vol. 142, no.~3, p. 031003, 2020.

\bibitem{ioannou2014l1}
P.~A. Ioannou, A.~M. Annaswamy, K.~S. Narendra, S.~Jafari, L.~Rudd, R.~Ortega,
  and J.~Boskovic, ``$\mathcal{L}_1$-adaptive control: {S}tability, robustness,
  and interpretations,'' \emph{IEEE Transactions on Automatic Control},
  vol.~59, no.~11, pp. 3075--3080, 2014.

\bibitem{pravitra2020MPPI}
J.~Pravitra, K.~A. Ackerman, C.~Cao, N.~Hovakimyan, and E.~A. Theodorou,
  ``$\mathcal{L}_1$-{Adaptive MPPI Architecture for Robust and Agile Control of
  Multirotors},'' in \emph{Proceedings of IEEE/RSJ International Conference on
  Intelligent Robots and Systems}.\hskip 1em plus 0.5em minus 0.4em\relax Las
  Vegas, NV, USA: IEEE, 2020, pp. 7661--7666.

\bibitem{diebel2006representing}
J.~Diebel, ``Representing attitude: {E}uler angles, unit quaternions, and
  rotation vectors,'' \emph{Matrix}, vol.~58, no. 15-16, pp. 1--35, 2006.

\bibitem{van1998real}
M.~J. Van~Nieuwstadt and R.~M. Murray, ``Real-time trajectory generation for
  differentially flat systems,'' \emph{International Journal of Robust and
  Nonlinear Control}, vol.~8, no.~11, pp. 995--1020, 1998.

\bibitem{lee2010geometricarxiv}
T.~Lee, M.~Leok, and N.~H. McClamroch, ``Control of complex maneuvers for
  quadrotor {UAV} using geometric methos on {SE} (3),'' \emph{arXiv:1003.2005},
  2010.

\bibitem{li2009l1}
D.~Li, N.~Hovakimyan, and C.~Cao, ``$\mathcal{L}_1$ adaptive controller in the
  presence of input saturation,'' in \emph{Proceedings of AIAA Guidance,
  Navigation, and Control Conference}, Chicago, IL, USA, 2009, p. 6064.

\bibitem{boyd1986note}
S.~Boyd, ``A note on parametric and nonparametric uncertainties in control
  systems,'' in \emph{Proceedings of American Control Conference}.\hskip 1em
  plus 0.5em minus 0.4em\relax Seattle, WA, USA: IEEE, 1986, pp. 1847--1849.

\bibitem{lakshmanan2020safe}
A.~Lakshmanan, A.~Gahlawat, and N.~Hovakimyan, ``Safe feedback motion planning:
  A contraction theory and $\mathcal{L}_1$-adaptive control based approach,''
  in \emph{Proceedings of the 59th IEEE Conference on Decision and
  Control}.\hskip 1em plus 0.5em minus 0.4em\relax Jeju, Korea: IEEE, 2020, pp.
  1578--1583.

\bibitem{MamboSimulink}
\BIBentryALTinterwordspacing
MathWorks. {Simulink Support Package for Parrot Minidrones}. [Online].
  Available: \url{https://www.mathworks.com/help/supportpkg/parrot/}
\BIBentrySTDinterwordspacing

\bibitem{berkenkamp2016safe}
F.~Berkenkamp, A.~P. Schoellig, and A.~Krause, ``Safe controller optimization
  for quadrotors with {G}aussian processes,'' in \emph{Proceedings of IEEE
  International Conference on Robotics and Automation}.\hskip 1em plus 0.5em
  minus 0.4em\relax Stockholm, Sweden: IEEE, 2016, pp. 491--496.

\bibitem{zhao2021RALPV}
P.~Zhao, S.~Snyder, N.~Hovakimyana, and C.~Cao, ``Robust adaptive control of
  linear parameter-varying systems with unmatched uncertainties,''
  \emph{arXiv:2010.04600}, 2021.

\bibitem{gahlawat2021contraction}
A.~Gahlawat, A.~Lakshmanan, L.~Song, A.~Patterson, Z.~Wu, N.~Hovakimyan, and
  E.~A. Theodorou, ``Contraction $\mathcal{L}_1$-adaptive control using
  {G}aussian processes,'' in \emph{Proceedings of the 3rd Learning for Dynamics
  \& Control}.\hskip 1em plus 0.5em minus 0.4em\relax Virtual Conference: PMLR,
  2021, pp. 1027--1040.

\end{thebibliography}
\end{document}